\documentclass[5p]{elsarticle}
\usepackage{graphicx}
\usepackage{amssymb}
\journal{Physics Letters A}

\usepackage{color}

\begin{document}

\begin{frontmatter}

\title{Strange Attractor in the Potts Spin Glass on Hierarchical Lattices \tnoteref{agencies}}
\author[ufpe2]{Washington de Lima}
\address[ufpe2]{Universidade Federal de Pernambuco, Centro Acadêmico do Agreste, Pernambuco, Brazil.}
\author[ufal]{G.\ Camelo-Neto\fnref{gcn}}
\address[ufal]{Universidade Federal de Alagoas,
N\'ucleo de Ci\^encias Exatas, 
Laborat\'orio de F\'{\i}sica Te\'orica e Computacional,
CEP 57309-005, Arapiraca, Alagoas, Brazil.}
\author[ufpe]{S.\ Coutinho\corref{sc}}
\ead{sergio@ufpe.br}
\address[ufpe]{Universidade Federal de Pernambuco, Departamento de F\'{\i}sica,
Laborat\'orio de F\'{\i}sica Te\'orica e Computacional,
Cidade Universit\'aria, CEP 50670-901, Recife, Pernambuco, Brazil.}
\tnotetext[agencies]{Work supported by the Brazilian granting agencies CNPq (grant PRONEX/FACEPE 0203-1.05/08), and CAPES}
\fntext[gcn]{Permanent Address: Universidade Federal de Pernambuco, Centro Acadêmico do Agreste, Pernambuco, Brazil.}
\cortext[sc]{Corresponding author: S. Coutinho}

\begin{abstract} The spin-glass $q$-state Potts model on $d$-dimensional diamond hierarchical lattices is investigated by an exact real space renormalization group scheme. Above a critical dimension $d_l(q)$ for $q > 2$, the coupling constants probability distribution flows to a low-temperature \emph{strange attractor} or to the high-temperature paramagnetic fixed point, according to the temperature is below or above the critical temperature $T_c(q,d)$. The strange attractor was investigated considering four initial different distributions for $q = 3$ and $d = 5$ presenting strong robustness in shape and temperature interval suggesting a condensed phase with algebraic decay.
\end{abstract}

\end{frontmatter}
\noindent Corresponding author:\\
Telephone: + 5581 21267633\\
Fax number: +5581 21268450 ext 2326\\
e-mail address: sergio@ufpe.br
\section{Introduction}

The $q$-states Potts model, proposed  a long time ago by Domb as the subject of Potts doctoral thesis~\cite{potts52}, has found a wide range of applicability in many fields of both basic and material sciences. The Potts model was conceived as a generalization of the Ising model~\cite{baxter82}, when $q = 2$, and the Askin-Teller model ($q = 4$)~\cite{ashkin43}. It also mimics the problem of percolation ($q = 1$)~\cite{broadbent57,kasteleyn69} and even the problem of the linear resistor networks ($q = 0$)\cite{wu82}. All of the above mentioned problems were also encompassed by the random-cluster model introduced by Fortuin and Kasteleyn~\cite{fortuin72a}. Furthermore, the degeneracy of the ground state of the \emph{antiferromagnetic} Potts models was shown to be related with the \emph{q-coloring} problem~\cite{fortuin72a}. It is important to emphasize that the most important feature of the mathematical structure of the Potts model is the equivalence between its partition function and Tutte polynomial~\cite{
tutte67}. Concerning applications, the Potts model has been applied in many fields, such as biology~\cite{sun04}, sociology~\cite{schulze05} and material science~\cite{miodownik07}. In the latter, for instance, the technique of Monte Carlo simulations on the Potts model has been applied to a wide variety of phenomena, such as diffusion in polycrystalline microstructures~\cite{swiler97} and the study of viscous instabilities in foam-flow behavior~\cite{sanyal06}. 

In this Letter, the properties of the $q$-state Potts model with random competing interactions are investigated. This model is called Potts glass in allusion to the particular case when $q = 2$, widely known in the literature as the Ising spin-glass model. The absence of spin-inversion symmetry and a different nature of the frustration effects distinguish it from its Ising counterpart, exhibiting rather a richer critical behavior in mean-field theory~\cite{gross85,cwilich89,cwilich90,gribova10, jani11a, jani11b,caltagirone12}, in contrast with the pure and disordered Potts model (without frustration) and related models, which have been widely investigated in the past \cite{wu82,tsallis96}. Site and bond diluted versions of the Ferro and Antiferromagnetic Potts model were also studied by Monte Carlo simulations on two and three-dimensional regular lattices~\cite{Chatelain2005,Murtazaev2012,Murtazaev2013} showing signatures of first and second order phase transitions.

More recently, the nature of phase transitions in the $q$-state Potts-glass model has been investigated via Monte Carlo simulation in two and three dimensions for values of the number of states $q = 3$ \cite{lee06}, $q=4$ \cite{cruz09}, $q = 5$ and $6$ \cite{banos10}, $q = 7$ \cite{oliveira09} and $q = 10$ \cite{lee06,brangian03}.

The present work focuses on the study of the $q$-state Potts model with random frustrated exchange interactions on a family of diamond-type hierarchical lattices~\cite{griffiths82} with scale factor $b=2$. When symmetrical zero-centered random exchange couplings distributions are considered the system undergoes a phase transition from a paramagnetic  high-temperature phase to a low-temperature condensed phase above some dimension, $d_l(q)$.  The phase diagram of the Potts-glass model had been investigated before, by the Migdal-Kadanoff (MK) renormalization group (RG) scheme, indicating the presence of a condensed phase at finite temperatures when $q > 2$ and $d = 4$ \cite{benyoussef96} as well as in the limit of large $q$~\cite{igloi09}.  In reference\cite{benyoussef96}, however, it was assumed (as working \emph{hypothesis}) that the initial symmetrical Gaussian probability distribution of coupling constants with variance $\sigma$ is transformed under renormalization into another symmetrical Gaussian distribution but with variance $\sqrt{b^{d-1}\sigma}$.

The MK real-space RG method for Bravais lattices is known to be equivalent to exactly solving the model on diamond-like hierarchical lattices\cite{kadanoff76,migdal76}. In a recent paper \cite{cameloneto04}, however, the authors studied the occurrence of phase transitions of the Potts-glass model using this exact approach by numerically following the flow of the renormalized probability distribution in an appropriated parameter space. Such space was previously considered in reference \cite{curado88,nogueira99} to study the Ising spin-glass model. The q-state Potts-glass model was considered on lattices with several fractal dimension, determining the critical temperature and the upper and lower bounds for the associated lower critical dimension $d_l(q)$.  For instance, for $q=3$ the lower (upper) bound was found to be 4.46 (4.58), in contrast with the result obtained in reference\cite{benyoussef96}, which founds the transition occurring for $d<4$. Here we further explore the flow of the renormalized probability distribution in the whole parameter space and investigate the nature of the low-temperature stable fixed point, which surprisingly appeared like a \emph{strange attractor}.

\section{Renormalization Procedure in Disordered Systems}

For  pure systems, the renormalization procedure consists in finding the equivalent exchange interaction for a pair of spins after eliminating several spins in the lattice. For a disordered system, however, the renormalization procedure will affect the whole distribution of coupling constants, the renormalized distribution, $P^\prime(J)$, is related with the previous (non-renormalized) distribution, $P(J)$, by,
\begin{equation}
P^\prime(J) = \int\cdots\int \prod_{<ij>}P(J_{ij})dJ_{ij}\delta(K-K^\prime(K))\mbox{,}
 \label{eq:renormalization_distribution}
\end{equation}
where $K=\beta J$ is a reduced coupling constant, $K^\prime(K)$ is the renormalization equation, and the product runs over all the pairs of spins $<ij>$.

Equation (\ref{eq:renormalization_distribution}) should be iterated until the renormalized distribution reaches a fixed point distribution, characteristic of the thermodynamic phase. A zero-centered Dirac-delta distribution, for instance, indicates a paramagnetic phase. The procedure adopted in this work is to produce a sample of random coupling constants from an initial probability density function, feed the renormalization equation to find a sample of the same size of renormalized coupling constants, estimating the physical quantities numerically from the samples. The process is repeated until the fixed point distribution is reached.

\section{The Potts Hamiltonian and the Renormalization Equation}

\begin{figure}
\begin{center}
\includegraphics*[width=7cm]{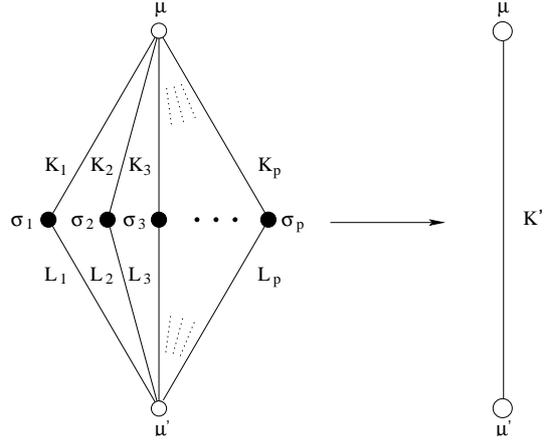}
\caption{Renormalization group scheme on the $d_f$ dimension diamond-type hierarchical lattice with $p$ branches and scaling factor $b=2$.}
\label{fig:diamante}
\end{center}
\end{figure}

The Potts Hamiltonian is written as
\begin{equation}
\mathcal{H} = -\sum_{<ij>}q\,J_{ij}\delta_{\sigma_i\,\sigma_j}\mbox{,}
\label{eq:Potts_hamiltonian}
\end{equation}
where the sum is taken over all bonds in the lattice, the Kronecker $\delta$ symbol takes the values $1$ if $\sigma_i=\sigma_j$ or $0$ otherwise, and $\sigma_i=1, 2,\ldots, q$ are the $q$-states Potts spins variables, located  at each site of a diamond-type hierarchical lattice with $p$ branches and scale factor $b$. The basic unit of such lattice is illustrated in figure~\ref{fig:diamante}, where $\mu$ and $\mu^\prime$ are called external sites and the set $\{\sigma_i\}$ represents the internal sites~\cite{griffiths82}. $K_i$ and $L_i$ are reduced coupling constants, $K_i\equiv \beta J_{\mu\sigma_i}$ and $L_i\equiv \beta J_{\mu'\sigma_i}$. The lattice generations or hierarchies are successively built by replacing each connection of the basic unit by the basic unit itself, yielding to a graph with fractal dimension,
\[
  d_f =1+ \frac{\ln{p}}{\ln{2}}\mbox{.}
\]

The exact renormalization process on a $n$--generation lattice consists in partially tracing the partition function along all the internal sites introduced in the $n^{th}$ generation leading to a ($n$-1)--generation lattice with a set of effective reduced coupling constants $\{K_i^\prime\}$ given by
\begin{equation}
\label{eq:renormalization_equation}
K^\prime = \frac{1}{q}\,\sum_{i=1}^p\left[\frac{(q-1) + \exp{(q\,K_i + q\,L_i)}}{(q-2) + \exp{(q\,K_i)} + \exp{(q\,L_i)}}\right]\mbox{.}
\end{equation}
Equation~(\ref{eq:renormalization_equation}) is the local renormalization equation. The right-hand side of equation (\ref{eq:renormalization_equation}) receives values for the reduced coupling constants calculated from the $n$--generation of the coupling constant distribution, $P(J)$, resulting in one of the possible values of the renormalized reduced coupling constant of the ($n-1$)--generation. The renormalization procedure starts from the thermodynamic limit (generation $n\rightarrow\infty$) where the coupling constants  are assumed to have a well-known distribution, actually, Gaussian, delta-bimodal, uniform, or exponential, and well-defined temperature $T$.

\subsection{Probability distribution renormalization flow}

For each $n$-generation lattice we can define a set of {\em thermal transmissivities} variables $\{t_{ij}\}$, each one associated with the respective bond, ie.,
\begin{equation}
t_{ij} \equiv \frac{1-\exp{(-q\beta J_{ij})}}{1+(q-1)\exp{(-q\beta J_{ij})}}\mbox{.}
\end{equation}
Thermal transmissivity $t_{ij}$ represents the pair correlation function $\Gamma_{ij}$ between sites $ij$~\cite{tsallis96}.

A system with a probability distribution of coupling constants $P(J)\equiv P(\{J_{ij}\})$ yields a thermal transmissivity variance,
\begin{equation}
\Delta^2 \equiv [(t_{ij}-[t_{ij}]_{J})^2]_{J}\mbox{,}
\end{equation}
where $[\cdots]_{J}$ means the average over the probability distribution, $P(J)$.

\begin{figure}[htbp]
\begin{center}
\includegraphics*[width=0.9\columnwidth]{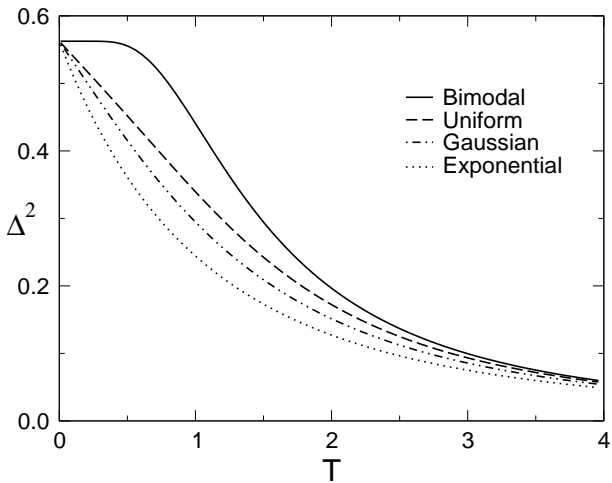}
\caption{$\Delta^2$ behavior for distinct probabilities distributions against temperature for the $q=3$ case.}
\label{fig:distributions}
\end{center}
\end{figure}

Every fixed probability distribution has a signature in the diagram $\Delta^2 \times T$ when the temperature is varied. In the case of the initial probability distributions considered in this work such representation can be seen in figure~\ref{fig:distributions} for the case $q=3$. 
Notice that the $T \to 0$ limit is the same for all cases indicating an universal zero-temperature behavior. Such universal point $(\Delta^{\star 2},T=0)$ can be analytically obtained in the case of the bimodal distribution, $P(J_{ij})=[\delta(J_{ij}-1)+\delta(J_{ij}+1)] /2$, ie.
\begin{equation}
\Delta^{\star 2}=\left[\frac{q}{2(q-1)}\right]^2\mbox{.}
\end{equation}
Note that for $q=2$, the Ising limit case, $\Delta^\star \to 1$  is recovered as seen in \cite{curado88}.

For each application of the renormalization equation, an effective temperature proportional to the inverse of the square root of the reduced coupling variance may be defined, ie.
\begin{equation}
T_r \propto \frac{1}{\sqrt{[(K_{ij}-[K_{ij}]_{J})^2]_{J}}}\mbox{.}
\end{equation}

In this work, the flow of the renormalized probability distributions is followed numerically in the parameter space $\Delta^2 \times T_r$, each point representing a stage of the renormalization process.

\subsection{Strange attractor}

Four distinct initial symmetrical probability distribution were considered, namely the Gaussian, the delta-bimodal, the uniform and the exponential ones.  For lattices with dimension greater than $d_l(q)$ and independently of the nature of the initial distribution, the renormalization flow displays the following same features: the paramagnetic phase is characterized by a \emph{fixed point} at infinity renormalized temperature and zero transmissivity variance, while the low-temperature phase is characterized by a \emph{strange attractor} in the renormalization flux for $q \geq 3$ -- or a fixed point $\Delta=1, T_r=0$ for $q=2$ as previously reported~\cite{curado88} for the Ising SG. The observation of such \emph{strange attractor}, which is located in a region of low but finite temperatures, is the main result to be reported in the present work. 

Chaotic renormalization group trajectories were observed in the Ising SG model a long time ago~\cite{mckay82} in a distinct family of hierarchical lattices. However, the signature of the existence of a strange attractor in the Potts-glass model on the diamond family of hierarchical lattices was first reported by Banavar and Bray~\cite{banavar88}. These authors observed an unusual behavior in the renormalization of $[K_{ij}]_J$ and its standard deviation in the case $d = 5$ and scale factor $b =2$ (present case). In their words these quantities ``wanders  chaotically in a smallish region around $[K_{ij}]_J\sim 14$ ... (persisting) for thousands of iterations and happens for a wide range of starting temperatures''. Furthermore, they did not give a clear explanation of this behavior but speculated that the delicate balance between energy and entropy causes that wandering phase in $d=5$.

This study uses the same methodology used in references \cite{curado88,banavar88}, ie. the successive construction of large pools of renormalized constant couplings, calculated using equation (\ref{eq:renormalization_equation}) starting from an initial probability distribution and fixed temperature.

\begin{figure}[ht]
\centering
\includegraphics*[width=0.9\columnwidth]{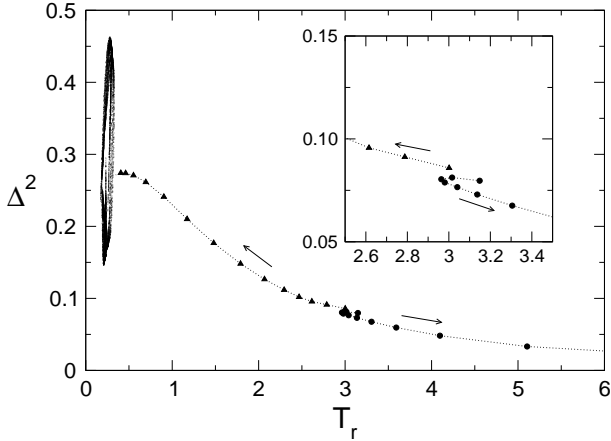}
\caption{Renormalization flow in the $\Delta^2\times T$ parameter space for the $q=3$-Potts model in $d=5$ dimensions for an initial symmetric Gaussian distribution. Dotted curve indicates the low-temperature flow while continuous one illustrates the high-temperature flow, according to the arrows. Inset magnifies the flow bifurcation region.}
\label{fig:renormalization_flow}
\end{figure}
\begin{figure}[ht]
\begin{center}
\includegraphics*[width=0.8\columnwidth]{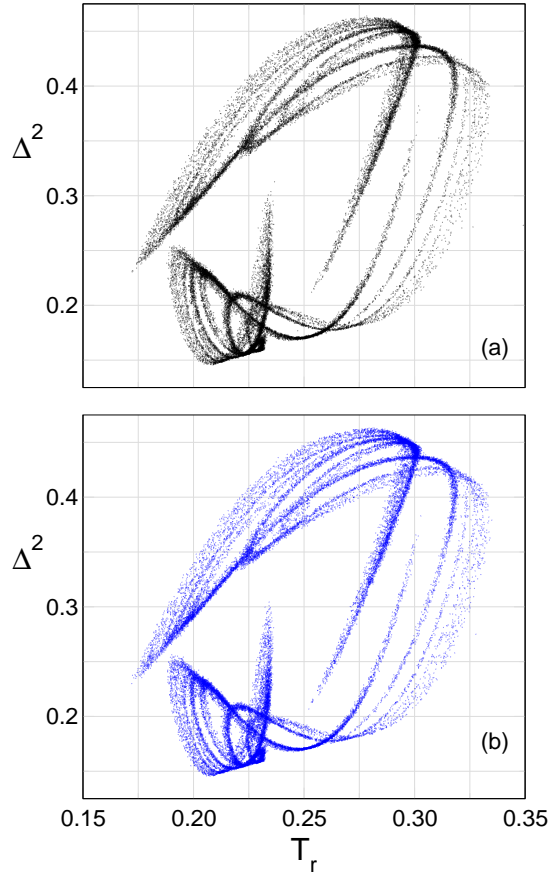}
\caption{Attractor for two distinct initial distribution, (a) Bimodal and (b) Gaussian.}
\label{fig:attractors}
\end{center}
\end{figure}

Figure~\ref{fig:renormalization_flow} illustrates the renormalization flow diagram in the $\Delta^2\times T$ space for the $q=3$ case in $d=5$ dimensions, starting from a Gaussian distribution of the exchange couplings in two temperatures. For initial temperatures below $\sim 3.0$ the renormalized distribution of interactions (solid triangles) flows to the strange attractor located in a finite region of the parameter space, for high initial temperatures ($T \sim 3.5$), the renormalized distribution of interactions (solid circles) flows to high temperatures and zero thermal transmissivity variance, characterizing the paramagnetic phase. Figure~\ref{fig:attractors} displays the strange attractor constructed with $10^5$ steps of renormalization, lying in the interval $(0.15, 0.35)$ starting from two distinct initial distributions, (a) the delta-bimodal and (b) the Gaussian ones.  

In reference\cite{cameloneto04}, some of the present authors, using the same methodology, found the critical temperatures of the model for several lattices, indicating the ones where the transition may occurs for appropriated values of $d$ and $q$.

The fractal dimension of the attractor was estimated by box counting, yielding to,
\[
 D = 1.66(1)\mbox{,}
\]
for every initial distribution and for the case $q=3$ and $d=5$. A proper analysis of the attractor is taking place at the moment, by estimating its Lyapunov exponents by the Wolf method~\cite{wolf85} with various values of $q$ and $d$.

The strange attractor describing the condensed phase occurs at a non-zero temperature interval probably due to the existence of residual entropy, such as the unusual algebraic phase previously studied~\cite{Berker1980,Qin1991,Redinz1994} for the anti-ferromagnetic Potts model. Detailed studies of the correlation functions and the local magnetization are under development in order to determine the nature of such condensed phase. Preliminary studies of the present model on lattices with scale factor $b=3$, which corresponds to bipartite lattices, indicates however the absence of the strange attractor in any dimension. Moreover, the condensed phase which occurs above a certain lower critical dimension $d_l (q)$ does not presents characteristics of spin-glass phase\cite{pedro12}.

\section{Conclusions}

The q-state spin-glass Potts model undergoes a phase transition whenever $d\geq d_l(q)$, from a high-temperature paramagnetic phase to a low-temperature condensed phase, characterized by strange attractors. These strange attractors are superimposed in the same region of parameter space independently of the initial distribution of coupling constants and initial temperatures below $T_c$, even if this initial temperature is chosen close to zero, below the attractor region. Such strange attractors, however, change in shape and position for distinct $(q, d)$ models in lattices with scale factor $b=2$. The occurrence of the strange attractor in a finite non-zero temperature interval suggests a condensed phase with algebraic decay of the correlation function.

\section{Acknowledgment}

We grateful acknowledge the enlightening and fruitful discussions and comments of E. M. F. Curado, W. A. M. Morgado, A. S. Rosas, and F. D. Nobre. This work received financial support from CNPq and CAPES (Brazilian federal grant agencies), from FACEPE (Pernambuco state grant agency under the grant PRONEX APQ 0203-1.05/08).

\end{document}